\documentclass[preprint,5p,times,twocolumn]{my-elsarticle}

\usepackage{amssymb}
\usepackage{amsmath}
\usepackage{mathrsfs}
\usepackage{mathtools}

\biboptions{compress}

\usepackage[usenames,table]{xcolor}
\usepackage[colorlinks=true,linkcolor=black,citecolor=blue,urlcolor=blue,filecolor=black]{hyperref}
\usepackage{comment}

\def\cH{{\cal H}}
\def\cI{{\cal I}}
\def\cJ{{\cal J}}

\def\cP{{\cal P}}

\def\be{\begin{equation}}
\def\ee{\end{equation}}

\begin{document}

\begin{frontmatter}



\title{Carrollian conformal scalar as flat-space singleton}


\author[Tours]{Xavier~Bekaert}
\author[UMONS]{Andrea~Campoleoni}
\author[UMONS,Edinburgh]{Simon~Pekar}

\address[Tours]{Institut Denis Poisson, Unit\'e Mixte de Recherche 7013, Universit\'e de Tours -- Universit\'e d’Orl\'eans -- CNRS,\\
Parc de Grandmont, 37200 Tours, France\\
{\tt\href{mailto:Xavier.Bekaert@lmpt.univ-tours.fr}{Xavier.Bekaert@lmpt.univ-tours.fr}}}

\address[UMONS]{Service de Physique de l'Univers, Champs et Gravitation, Universit\'e de Mons -- UMONS,\\ 20 place du Parc, 7000 Mons, Belgium\\
{\tt \href{mailto:andrea.campoleoni@umons.ac.be}{andrea.campoleoni@umons.ac.be}},\quad
{\tt \href{mailto:simon.pekar@umons.ac.be}{simon.pekar@umons.ac.be}}}

\address[Edinburgh]{School of Mathematics and Maxwell Institute, \\ University of Edinburgh, \\ Peter Guthrie Tait Road, Edinburgh EH9 3FD, UK}

\begin{abstract}
We show that, in any space-time dimension, the on-shell (electric) conformal Carrollian scalar can be interpreted as the flat-space limit of the singleton representation of the conformal algebra. 
In fact, a recently proposed higher-spin algebra for Minkowski spacetime amounts to the Poincar\'e enveloping algebra on the corresponding module.
This higher-spin algebra is a contraction of that entering Vasiliev's equations, which can be constructed analogously from the singleton representation of the conformal algebra.
We also show that the higher-spin extension of the Poincar\'e algebra we consider is a subalgebra of all symmetries of the conformal Carrollian scalar, given by a higher-spin version of the (extended) BMS algebra.
\end{abstract}

%\begin{keyword}



%\end{keyword}

\end{frontmatter}


%% main text %%%%%%%%%%%%%%%%%%%%%%%%%%%%%%%%%%%%%%%%%%%%%%%%%%%%%%

\section{Introduction}

Even if the holographic principle does not seem to require a negative cosmological constant, most of its concrete realisations ---~in which one has an independent definition of both sides of the duality~--- involve a gravitational theory with asymptotically Anti de Sitter (AdS) solutions. Recent years have seen  enormous progress towards a flat-space holographic correspondence (see, e.g., \cite{Pasterski:2021raf, Donnay:2022aba, Bagchi:2022emh} for an overview of various approaches), but only a handful of proposals involving specific boundary theories are available \cite{Dappiaggi:2005ci, Kar:2022vqy, Costello:2022jpg, Stieberger:2022zyk, Ponomarev:2022ryp}. It may thus be useful to build additional holographic dualities in flat spacetime, even involving other gravitational theories than general relativity. 

One of the simplest holographic dualities in AdS relates higher-spin gravity in the bulk with a free scalar field on the boundary, in any space-time dimensions (see, e.g., \cite{Bekaert:2012ux, Giombi:2016ejx} for a review). From the point of view of the algebra $\mathfrak{so}(2,d+1)$, a free conformal scalar on the boundary of AdS$_{d+2}$ corresponds to Dirac's singleton representation \cite{Dirac:1963ta}. The tensor product of two singletons \cite{Flato:1978qz, Vasiliev:2004cm} then reproduces the spectrum of Vasiliev's bosonic equations \cite{Vasiliev:2003ev}, thus guaranteeing that the infinite tower of conserved currents of the boundary theory can couple to the bulk higher-spin fields. This result somehow anticipated higher-spin holography and lies at its foundations: the Lie algebra ruling Vasiliev's equations, that we denote by $\mathfrak{hs}_{d+2}$, coincides with the universal enveloping algebra (UEA) of $\mathfrak{so}(2,d+1)$ realised on the singleton module \cite{Vasiliev:1999ba, Eastwood:2002su, Iazeolla:2008ix}. Equivalently, one can obtain the algebra $\mathfrak{hs}_{d+2}$, often called Eastwood-Vasiliev algebra, by quotienting the $\mathfrak{so}(2,d+1)$ UEA by an ideal that identically vanishes on the singleton module.

In sections~\ref{sec:algebra} and \ref{sec:ambient}, we exhibit a similar algebraic setup for $(d+2)$-dimensional Minkowski spacetime: we identify a (non-unitary) representation of the Poincar\'e algebra $\mathfrak{iso}(1,d+1)$, that we dub \emph{simpleton},\footnote{This nickname originates from two facts: the simpleton is the flat-space analogue of the \textit{singleton}, as shown here, and it corresponds to one of the \textit{simplest} possible Carrollian field theories, cf.\ the action \eqref{eq:action}.} and we show that 
\begin{enumerate}
\item the flat-space higher-spin algebra of \cite{Campoleoni:2021blr}, which is a contraction of the Eastwood-Vasiliev algebra, corresponds to the Poincar\'e UEA realised on the simpleton module;
\item the simpleton admits a realisation as an on-shell conformal Carrollian scalar field on null infinity, in its ``electric'' or ``time-like'' version.
\end{enumerate}
The latter is a scalar field living on $\mathscr{I}\cong\mathbb{R}\times S^d$, with scaling dimension $\Delta = \frac{d-1}{2}$ and with action
\be \label{eq:action}
S[\varphi] = \frac12 \int du\, d^d x \sqrt{\gamma}\,\, \partial_u\,\varphi^*\,\partial_u\,\varphi \,,
\ee
where $\gamma_{ij}$ is the metric on the $d$-dimensional unit sphere, while $u$ is the (retarded) time. 
This action results from the massless Klein-Gordon one in the limit in which the speed of light tends to zero \cite{Bagchi:2019xfx, Henneaux:2021yzg, deBoer:2021jej} and it can be generalised to curved Carrollian manifolds \cite{Gupta:2020dtl, Rivera-Betancour:2022lkc, Baiguera:2022lsw}. In the $c \to 0$ limit, the conformal symmetry of a free scalar in $d+1$ dimensions contracts to the conformal Carroll symmetry, which is isomorphic to the Poincar\'e one in $d+2$ dimensions \cite{Duval:2014uva}. Roughly speaking, in the following we show that similar considerations apply to the additional, higher-spin symmetries of the free relativistic scalar \cite{Nikitin1991-gp, Eastwood:2002su}.

Given the crucial role played by symmetries in higher-spin holography, it is tempting to consider these results as an indication that the conformal Carrollian field theory \eqref{eq:action} may enter a holographic duality involving higher-spin fields in Minkowski spacetime, thus providing a simple realisation of ``Carrollian holography'' along the lines of, e.g., \cite{Arcioni:2003td, Bagchi:2016bcd, Ciambelli:2018wre, Donnay:2022aba, Bagchi:2022emh}.
The boundary theory would be among the simplest examples of conformal Carrollian field theories and, in spite of all no-go theorems, evidence is accumulating that higher-spin theories in flat spacetime can be defined at the price of introducing some unconventional features. For instance, concrete steps towards a flat-space avatar of higher-spin holography were performed in \cite{Ponomarev:2022ryp, Ponomarev:2022qkx} for chiral higher-spin gravity \cite{Ponomarev:2016lrm}. Still, this model involves a different higher-spin algebra compared to \cite{Campoleoni:2021blr} and, therefore, it leads to an alternative flat-space analogue of the singleton.

Other (bosonic) higher-spin algebras for Minkowski space were also presented in \cite{Bekaert:2008sa, Campoleoni:2021blr, Bekaert:2022ipg}, together with infinite-dimensional extensions \cite{Bekaert:2022ipg} in the spirit of the Bondi-Metzner-Sachs (BMS) extension of the Poincar\'e algebra. BMS-like higher-spin symmetries in any dimensions already appeared as  asymptotic symmetries for fields of arbitrary spin in Minkowski space \cite{Campoleoni:2017mbt, Campoleoni:2020ejn} and we recover them too in section~\ref{sec:symmetries}. There, we classify all symmetries of the simpleton and obtain that the algebra of \cite{Campoleoni:2021blr} does not capture all of them, in sharp contrast from what happens for a relativistic scalar \cite{Nikitin1991-gp, Eastwood:2002su}. 

\section{Carrollian scalar and higher-spin algebra} \label{sec:algebra}

We begin by reviewing the construction of the flat-space higher-spin algebra $\mathfrak{ihs}_{d+2}$ of \cite{Campoleoni:2021blr}. 
Consider the Poincar\'e algebra $\mathfrak{iso}(1,d+1)$ of isometries of $(d+2)$-dimensional Minkowski space, spanned by translation generators $\cP_a$ and Lorentz generators $\cJ_{bc}$, with $a,b,c \in \{0,1,\dots,d+1\}$:
\begin{subequations}
\begin{align}
[\cJ_{ab}, \cJ_{cd}] &= \eta_{ac}\,\cJ_{bd} - \eta_{ad}\,\cJ_{bc} - \eta_{bc}\,\cJ_{ad} + \eta_{bd}\,\cJ_{ac} \,,\\
[\cJ_{ab}, \cP_c] &= \eta_{ac}\,\cP_b - \eta_{bc}\,\cP_a \,,\\
[\cP_a, \cP_b] &= 0 \,.
\end{align}
\end{subequations}
For $d \geq 2$, the Lie algebra $\mathfrak{ihs}_{d+2}$ is the UEA of $\mathfrak{iso}(1,d+1)$ quotiented by the (two-sided) ideal spanned by\footnote{A similar construction can be employed also when $d=1$, although \eqref{eq:ideal-antisymmetric-2} is trivial, and it corresponds to the $\lambda = 1$ case of the $\mathfrak{ihs}_3[\lambda]$ one-parameter family of non-isomorphic higher-spin algebras \cite{Ammon:2020fxs, Campoleoni:2021blr}.
}
\begin{subequations} \label{eq:ideal}
\begin{align}
\label{eq:ideal-translations} \{ \cP_a,\, \cP_b \} &\sim 0 \,,\\[4pt]
\label{eq:ideal-translation-Lorentz} \cI_a \equiv \{\cP^b,\, \cJ_{ba} \} &\sim 0 \,,\\[5pt]
\label{eq:ideal-antisymmetric-1} \cI_{abc} \equiv \{\cP_{[a},\, \cJ_{bc]} \} &\sim 0 \,,\\[5pt]
\label{eq:ideal-antisymmetric-2} \cI_{abcd} \equiv \{\cJ_{[ab},\, \cJ_{cd]} \} &\sim 0 \,,\\
\label{eq:ideal-eigenvalue} \cJ^2 + \frac{d^2-1}{4}
&\sim 0 \,.
\end{align}
\end{subequations}
Here and below, the weak equality symbol $\sim$ stands for ``equal modulo terms in the ideal''. The associative product on the UEA will be implied in the following and $\{a,b\} = a\, b + b\, a$. The relation \eqref{eq:ideal-eigenvalue} fixes the eigenvalue of the quadratic Casimir operator of the Lorentz subalgebra \mbox{$\mathfrak{so}(1,d+1)$}:
\be
\cJ^2 \equiv \frac12\,\cJ_{ab}\,\cJ^{ba} \,.
\ee
This identification is possible even if $\cJ^2$ is not a Casimir operator for the full Poincar\'e algebra thanks to the other relations in \eqref{eq:ideal}. The above ideal results from the flat-space limit \cite{Campoleoni:2021blr} of the ideal one has to quotient out from the $\mathfrak{so}(2,d+1)$ UEA to obtain the Eastwood-Vasiliev algebra $\mathfrak{hs}_{d+2}$. In its turn, the latter ideal is the annihilator of the singleton module, i.e., it vanishes when the $\mathfrak{so}(2,d+1)$ UEA is evaluated on this module \cite{Vasiliev:1999ba, Eastwood:2002su, Iazeolla:2008ix}.

Note that the last two relations in the ideal \eqref{eq:ideal} imply the first three ones. In fact, eqs.~\eqref{eq:ideal-translation-Lorentz} and \eqref{eq:ideal-translations} follow successively from \eqref{eq:ideal-eigenvalue} by considering the adjoint action of translations twice, while \eqref{eq:ideal-antisymmetric-1} follows from \eqref{eq:ideal-antisymmetric-2} by considering the adjoint action of a translation. Therefore, the full ideal is actually generated by the conditions
\be \label{eq:ideal-small}
\cI_{abcd} \sim 0 \,, \quad \cJ^2 + \frac{d^2-1}{4}
\sim 0
\ee
involving only the Lorentz subalgebra, which thus suffice to give an implicit definition of the ``simpleton'' representation. Conversely, the eigenvalue of the Lorentz quadratic Casimir in \eqref{eq:ideal-eigenvalue} is fixed by
\be
\begin{split}
&\cI_{abc}\,\cJ^{bc}\! + \frac23\,\cJ_{ab}\,\cI^b\! + \frac{d-1}{3}\,\cI_a  = -\frac43 \left[ \cJ^2 + \frac{d^2-1}{4}
\right]\cP_a \,,
\end{split}
\ee
implying that factorising $\cI_{abc}$ and $\cI_a$ while keeping $\cP_a$ non-zero in the enveloping algebra requires \eqref{eq:ideal-eigenvalue}. The relations \eqref{eq:ideal-small} identify a subalgebra made only of products of $\cJ_{ab}$ of the type considered in \cite{Joung:2015jza}, and the rest of the algebra $\mathfrak{ihs}_{d+2}$ is obtained by the successive adjoint action of $\cP_a$ on it.
Clearly, one can instead add on top of \eqref{eq:ideal} the condition that translations act trivially, that is $\cP_a \sim 0$, as in the flat-limit of the singleton proposed in \cite{Flato:1978qz}. Still, representations satisfying \eqref{eq:ideal} with a non-trivial, although nilpotent, action of translations exist. More precisely, a representation with these properties is unique for $d>1$ and coincides with the space of solutions of \eqref{eq:action}.\footnote{In three and four space-time dimensions, such representations with non-trivial translations and satisfying \eqref{eq:ideal} were already exhibited in \cite{Campoleoni:2021blr, Ponomarev:2021xdq}, although the identification with the space of solutions of \eqref{eq:action} was missing. When $d=1$, one can also relax the condition \eqref{eq:ideal-eigenvalue} and obtain a one-parameter family of representations with non-trivial translations satisfying \eqref{eq:ideal-translations} \cite{Campoleoni:2021blr}.}

We now show that, indeed, the relations \eqref{eq:ideal} are identically satisfied on the module defined by the on-shell (electric) conformal Carrollian scalar field \eqref{eq:action} defined on $\mathscr{I}\cong\mathbb{R}\times S^d$.
To parameterise this space, we introduce the retarded time $u$ and the coordinate angles $x^i$ on the celestial sphere, where $i,j\in\{1,2,\dots,d\}$.
As shown in section~\ref{sec:ambient}, a representation of the Poincar\'e algebra is given by the following differential operators, acting on a scalar field $\varphi(u,x)$ of arbitrary scaling dimension $\Delta$ as
\begin{subequations} \label{realisation}
\begin{align}
\cP_a & = f_a(x)\,\partial_u \,, \label{realisation_P}\\
\cJ_{ab} & = \xi_{[ab]}^i(x) \,\partial_{i} + \frac{1}{d}\,\nabla_{\!i}\,\xi_{[ab]}^i(x) \left(\Delta + u\,\partial_u\right) , \label{realisation_J}
\end{align}
\end{subequations}
where $\nabla$ is the Levi-Civita connection for the metric $\gamma_{ij}$ on the unit sphere $S^d$. Moreover, the $f_a$'s are the $d+2$ solutions to the ``good-cut'' equation (see, e.g., \cite{Herfray:2021qmp} for a review)
\be\label{goodcut}
\nabla_{\!(i} \nabla_{\!j)}\,f_a = \frac{1}{d}\,\gamma_{ij}\,\nabla^2 f_a \,,
\ee
and the $\xi^{[ab]}$'s are the $\frac{(d+1)(d+2)}{2}$ conformal Killing vector fields of the celestial sphere $S^d$, that solve the equation\footnote{Relaxing the condition \eqref{goodcut}, the generators \eqref{realisation} define a representation of the BMS algebra.
The extended BMS algebra of \cite{Campiglia:2014yka} is obtained ignoring also \eqref{conformalKilling}.} 
\be\label{conformalKilling}
\nabla_{\!(i}\,\xi_{j)}{}^{[ab]}
= \frac{1}{d}\,\gamma_{ij}\,\nabla\cdot \xi^{[ab]} \,.
\ee
The previous realisation of the Poincar\'e  algebra satisfies the conditions \eqref{eq:ideal-antisymmetric-1} and \eqref{eq:ideal-antisymmetric-2} for any value $\Delta$ of the scaling dimension, as we prove with ambient-space techniques in section~\ref{sec:ambient}. 

It is important to note that the operators $\cJ_{ab}$ in \eqref{realisation} correspond to the standard realisation of the generators of the conformal algebra $\mathfrak{so}(1,d+1)$ acting on a  primary scalar field on the sphere $S^d$ with dimension $\Delta$, up to the replacement of the number $\Delta$ by the first-order operator $\Delta+u\,\partial_u$ (thereby taking into account the scaling property of $u$). 
Therefore, when acting on a ``Carrollian primary'' scalar $\varphi$ on $\mathscr{I}$ with scaling dimension $\Delta$, we have
\be\label{Jsquared}
\cJ^2\,\varphi 
=\left(\Delta + u\,\partial_u\right)\left(\Delta + u\,\partial_u - d\right) \varphi \,,
\ee
which originates from the usual result $\cJ^2\,\varphi =\Delta(\Delta - d)\,\varphi$ for a conformal primary scalar of dimension $\Delta$, taking into account the shift $\Delta \to \Delta + u\,\partial_u$. Eq.~\eqref{Jsquared} can also be checked directly using the ambient-space techniques of section~\ref{sec:ambient}.

We now assume that the condition \eqref{eq:ideal-eigenvalue} on the eigenvalue of $\cJ^2$ is satisfied as well. This hypothesis implies
\be\label{Carrollconfscalar}
\left[u^2{\partial_u}^{\!2} + \left(2\Delta - d + 1\right) u\,\partial_u + \Delta \left(\Delta - d\right)+\frac{d^2-1}4 \right] \varphi \sim 0 \,.
\ee
Let us stress that eq.~\eqref{eq:ideal-eigenvalue} is part of an ideal: in particular, this means that its commutator with any element of the Poincar\'e algebra must vanish as well. The adjoint action of $\mathcal P_0 = \partial_u$ on the differential operator entering \eqref{Carrollconfscalar} gives successively
\be
\left[2\,u\,{\partial_u}^2 + \left(2\,\Delta - d + 1\right) \partial_u\right]\varphi \sim 0 \,,\qquad {\partial_u}^2\,\varphi \sim 0 \,.
\ee
All previous conditions are satisfied if and only if
\be
{\partial_u}^2\,\varphi \sim 0 \,,\qquad \Delta = \frac{d-1}{2} \,,
\ee
which correspond to the equation of motion and to the scaling dimension of the conformal Carrollian scalar with action \eqref{eq:action}.

All in all, we discovered that the electric scalar field \eqref{eq:action} satisfies on shell the conditions \eqref{eq:ideal-small} and, therefore, the full set of conditions \eqref{eq:ideal}. As a result, the ideal of the Poincar\'e UEA that one has to factor out to obtain the algebra $\mathfrak{ihs}_{d+2}$ should coincide with the annihilator of the conformal Carrollian scalar.\footnote{Strictly speaking, one should also check that, in analogy with the case of a relativistic scalar, the whole annihilator of the conformal Carrollian scalar is generated by the relations \eqref{eq:ideal-small}, i.e., that there are no other independent combinations of the generators that vanish identically on the simpleton module.} Consequently, as we discuss in section~\ref{sec:symmetries}, symmetrised products of the generators \eqref{realisation} should give the whole $\mathfrak{ihs}_{d+2}$ algebra.

To conclude this section we mention that, as anticipated in \cite{Ponomarev:2021xdq, Ponomarev:2022ryp} and differently from \cite{Flato:1978qz, Vasiliev:2004cm}, the tensor product of two simpletons does not seem to decompose in representations of the Poincar\'e algebra corresponding to bulk higher-spin fields, because bilinears in the simpleton are at most quadratic 
in $u$ while the boundary data of radiative solutions are arbitrary functions of $u$. This feature may look disturbing in view of possible applications of our findings in an holographic setup, but we stress that in ``Carrollian holography'' radiation reaching null infinity might be included as an external source to which the boundary theory should couple \cite{Donnay:2022aba, Bagchi:2022emh}. This suggests some options to recover the d.o.f.\ of bulk fields via other mechanisms.

\section{Ambient space construction of the simpleton} \label{sec:ambient}

In this section, we prove that the representation \eqref{realisation} of the Poincar\'e algebra satisfies eqs.~\eqref{eq:ideal-small} by realising $\mathscr{I}$ as an embedded manifold in an ambient space and the simpleton as an on-shell ambient field. To this end, it is useful to first recall the ambient-space realisation of Dirac's singleton (see, e.g., \cite{Bekaert:2012vt} for more details). Consider the ambient space $\mathbb R^{d+1,2}$ with Cartesian coordinates $X^A$ and metric $\tilde \eta_{AB} = \text{diag}(-,+,\dots,+,-)$: the singleton is a field $\Phi(X^A)$ satisfying
\be \label{sp2}
\tilde \eta^{AB} \partial_A \partial_B \Phi = 0 \,,\quad \left(X^A\partial_A + \Delta\right) \Phi = 0 \,,\quad \Phi \simeq \Phi + X^2\,\Psi 
\ee
for any $\Psi(X^A)$, with $X^2 = \tilde \eta_{AB}\,X^A\,X^B$. The three scalar operators in \eqref{sp2} form the $\mathfrak{sp}(2)$ algebra provided that $\Delta = \frac{d-1}{2}$. In this framework, the free scalar field $\varphi$ living on the boundary of AdS$_{d+2}$ is recovered as the restriction of $\Phi$ to the light-cone \mbox{$X^2 = 0$}. The ambient-space isometries which preserve the light-cone are generated by
\be \label{eq:ambient-iso}
J_{AB} \equiv 2\,X_{[A} \partial_{B]} \,.
\ee
In this realisation, the ideal that one has to factor out in order to recover the higher-spin algebra $\mathfrak{hs}_{d+2}$ is automatically factorised for $\Delta = \frac{d-1}{2}$ since
\begin{subequations}\label{ideal-AdS}
\begin{align}
J_{[AB}\,J_{CD]}\,\Phi & = 4\,X_{[A}\,X_C\,\partial_B\,\partial_{D]}\,\Phi + 4\,\eta_{[BC}\,X_A\,\partial_{D]}\,\Phi = 0 \,, \\
J^C{}_{(A}\,J_{B)C}\,\Phi & \simeq - \tilde\eta_{AB}\,\Delta\,\Phi \,,\quad \\J^2\,\Phi &= \Delta\left(\Delta - 1 - d\right) \Phi \,.
\end{align}
\end{subequations}
The conditions \eqref{eq:ideal} we are interested in can then be recovered from \eqref{ideal-AdS} by sending the cosmological constant to zero \cite{Campoleoni:2021blr}.

On the other hand, they can also be obtained by embedding null infinity in an ambient space with a degenerate metric and following similar steps as before. The starting point is now the Carrollian ambient space $\mathbb R \times \mathbb R^{d+1,1}$ with coordinates $(u,y^a)$ \cite{Herfray:2020rvq}. We choose the degenerate ambient metric to coincide with the Minkowski metric $\eta_{ab}$ in \mbox{$d+2$} dimensions (so that the null coordinate $u$ parameterises its degenerate direction).
The simpleton is then defined as a field $\Phi(u,y^a)$ satisfying
\be \label{iso11}
{\partial_u}^2\,\Phi = 0 \,,\quad \left(y^a\partial_a + u\,\partial_u + \Delta\right)\Phi = 0 \,,\quad \Phi \simeq \Phi + y^2\,\Psi
\ee
for any $\Psi(u,y^a)$, together with the additional condition \mbox{$\Delta = \frac{d-1}{2}$}. Both the degenerate ambient-space metric and the constraints \eqref{iso11} can be obtained as a Carrollian, $c \to 0$ limit of the previous AdS ambient-space construction, thus justifying the additional constraint on the scaling dimension $\Delta$.\footnote{Since we are considering a $c \to 0$ limit of the relativistic ambient-space construction, geometrically we are actually realising $\mathscr{I}$ as the boundary of an AdS-Carroll manifold (see, e.g., \cite{Figueroa-OFarrill:2018ilb}). It will be interesting to use directly an ambient-space construction suited to Minkowski space \cite{Herfray:2020rvq}, but the current framework anyway suffices to prove the relations \eqref{eq:ideal} for the representation \eqref{realisation}.} Note, however, that  for any value of $\Delta$ the differential operators in \eqref{iso11} satisfy the algebra $\mathfrak{iso}(1,1)$, which is a contraction of $\mathfrak{sp}(2)\cong\mathfrak{so}(2,1)$. 

In this framework, $\mathscr{I}$ corresponds to the locus $y^2 = 0$ for any value of $u$ and the ambient-space isometries that preserve \mbox{$y^2 = 0$}, are analytic in the coordinates and have zero homogeneity degree, i.e.\ that are invariant under $(u,y^a) \to (\lambda\,u,\lambda\,y^a)$, read
\be \label{null-isometries}
\mathcal J_{ab} \equiv 2\,y_{[a}\,\partial_{b]} \,,\quad \mathcal P_a \equiv y_a\,\partial_u \,.
\ee
They can also be obtained as the $c \to 0$ limit of the $\mathfrak{so}(2,d+1)$ isometries in \eqref{eq:ambient-iso}. In the particular differential realisation of \eqref{null-isometries}, the relations \eqref{eq:ideal-antisymmetric-1} and \eqref{eq:ideal-antisymmetric-2} are automatically satisfied,
\be
\mathcal I_{abcd}\,\Phi = 0 \,,\quad \mathcal I_{abc}\,\Phi = 0 \,,
\ee
while eqs.~\eqref{eq:ideal-translations}, \eqref{eq:ideal-translation-Lorentz} and \eqref{eq:ideal-eigenvalue} are verified provided that the constraints \eqref{iso11} are satisfied and $\Delta = \frac{d-1}{2}$:
\be
\mathcal P_a\,\mathcal P_b\,\Phi = 0 \,,\quad \mathcal I_a\,\Phi \simeq 0 \,,\quad \left(\mathcal J^2 + \frac{d^2-1}{4}\right)\Phi \simeq 0 \, .
\ee

We now show that an ambient function $\Phi(u,y^a)$ obeying to the last two conditions in \eqref{iso11} defines a Carrollian conformal primary field $\varphi(u,x^i)$ with scaling dimension $\Delta$, upon restriction to the submanifold $y^2=0$; and that, as differential operators on $\mathscr{I} \cong \mathbb R \times S^d$, the operators in \eqref{null-isometries} correspond to
those in \eqref{realisation}.
Explicitly, one can write the non-degenerate part of the ambient metric in a null form using the coordinates $y^a = (y^0, y^i, y^\infty)$ so as to obtain
\be
\eta_{ab}\,y^a\,y^b = 2\,y^0\,y^\infty + \gamma_{ij}\,y^i\,y^j \,.
\ee
Then, the locus $y^2 = 0$ in the vicinity of $y^0 > 0$ can be parameterised by 
\be
y^0 = x^0\,, \quad y^i = x^0\,x^i\,, \quad y^\infty = -\frac12 x^0 x^2\,.
\ee
Since $\Phi$ is homogeneous of degree $\Delta$, i.e.\ $\Phi(\lambda\,u, \lambda\,y^a) = \lambda^{-\Delta} \Phi(u,y^a)$ for $\lambda > 0$, we can identify $(u,y^a)$ with $\frac{1}{x^0}(u,y^a)$ for all $x^0 > 0$. In practice, we can set $x^0 = 1$ in this region. Then, in this choice of coordinates, it is manifest that $y^a = (1,x^i,-\frac12 x^2)$ is the unique $\mathbb R^{d+1,1}$-vector of functions of $x^i$ verifying the ``good-cut'' equation \eqref{goodcut} (with $f^a \equiv y^a$). This proves that the $\cP_a$ in \eqref{null-isometries} correspond to those in \eqref{realisation}. Similarly, following the same argument as in \cite{Eastwood:2002su}, $\xi_i^{[ab]}$ is the unique rank-two antisymmetric tensor of functions of $x^i$ verifying eq.~\eqref{conformalKilling}.

Note that the $y^a(x^i)$ and $\xi_i^{[ab]}(x^i)$ satisfying eqs.~\eqref{goodcut} and \eqref{conformalKilling} can also be interpreted as generalised conformal Killing scalars and vectors with depths 1 and 0 respectively.
Following the discussion of generalised conformal Killing tensors of e.g.~\cite{Bekaert:2013zya}, one can then realise the good-cut and conformal Killing equations \eqref{goodcut} and \eqref{conformalKilling} in ambient space as the following pairs of equations:
\begin{subequations}
\begin{alignat}{5}
\partial_a\,\partial_b\,f(y) &= 0\,, & \qquad \left(y^a \partial_a - 1\right)f(y) &= 0\,, \\
\partial_{(a}\,\xi_{b)}(y) &= 0\,, & \qquad \left(y^a\partial_a - 1\right)\xi_b(y) &= 0\,.
\end{alignat}
\end{subequations}
These equations are clearly satisfied by the components of the isometry vectors \eqref{null-isometries}, thus confirming once again their identification with the vectors \eqref{realisation}.

\section{All symmetries of the simpleton}\label{sec:symmetries}

Following the philosophy of \cite{Eastwood:2002su} (see also \cite{Nikitin1991-gp, Shapovalov1992-tf}), we now show that a real form of the  higher-spin algebra $\mathfrak{ihs}_{d+2}$ is a subalgebra of the higher symmetries of the conformal Carrollian scalar. It was already noticed that the higher symmetries of \eqref{eq:action} contain all generators of the form $f(x)\,\partial_u$ without any constraints on the functions $f(x)$ \cite{Bagchi:2019xfx} (see also \cite{Bidussi:2021nmp, Hao:2021urq, Bagchi:2022eav}), so that they include, at least, super-translations.
Similarly, in classifying the higher symmetries of \eqref{eq:action}, we shall obtain infinite-dimensional extensions of the algebra $\mathfrak{ihs}_{d+2}$ incorporating (extended) BMS symmetries and BMS-like higher-spin symmetries similar to those in \cite{Campoleoni:2017mbt, Campoleoni:2020ejn, Bekaert:2022ipg}.

\subsection{Higher symmetries of the conformal Carrollian scalar}

We define a higher symmetry of the quadratic action \eqref{eq:action} to be a differential operator $D$ on $\mathscr{I}$ such that the infinitesimal transformation $\delta\varphi = i\epsilon D\varphi$ leaves \eqref{eq:action} invariant, or in other words, such that $D$ weakly commutes with the kinetic operator ${\partial_u}^2$ in the sense that
\be\label{hermiticity}
{\partial_u}^2 \circ D = D^\dagger \circ {\partial_u}^2 \,,
\ee
where $^\dagger$ is the Hermitian conjugation with respect to the inner product $\langle\,\psi\mid\varphi\,\rangle =  \int du\, d^d x \sqrt{\gamma}\,\psi^*\varphi$.
Since $\left[ {\partial_u}^2 , D \right] = 2\,\dot D \circ \partial_u + \ddot D$, the condition \eqref{hermiticity} is equivalent to
\be \label{eq:weak-commutator}
2\,\dot D \circ \partial_u + \ddot D = \left(D^\dagger - D \right) \circ {\partial_u}^2 \,.
\ee
By implementing the on-shell identification $D \sim D + B \circ {\partial_u}^2$ for any differential operator $B$, one can write down without loss of generality an ansatz for $D$ of the form
\be
D = D_0 + D_1 \circ \partial_u \,,
\ee
where $D_0$ and $D_1$ are independent of $\partial_u$. Then, eq.~\eqref{eq:weak-commutator} translates into
\be
\begin{split}
&2\,\left(\dot D_0 + \dot D_1 \circ \partial_u\right) \circ \partial_u + \left(\ddot D_0 + \ddot D_1 \circ \partial_u\right) \\
&\qquad = \left[ \left(D_0{}^\dagger - D_0\right) - \left(D_1{}^\dagger + D_1\right) \circ \partial_u - \dot D_1{}^\dagger \right] \circ \partial_u{}^2 \,,
\end{split}
\ee
which decomposes into powers of $\partial_u$ as
\begin{subequations}
\begin{alignat}{5}
D_1{}^\dagger + D_1 &= 0 \,, & \qquad
2\,\dot D_1 &= D_0{}^\dagger - D_0 - \dot D_1{}^\dagger \,,\\
2\,\dot D_0 + \ddot D_1 &= 0 \,, & \qquad
\ddot D_0 &= 0 \,.
\end{alignat}
\end{subequations}
The general solution is
\be
D_0 = K_0 - iK_{+1}\,u \,,\quad D_1 = iK_{-1} + \left(K_0{}^\dagger - K_0\right)u + iK_{+1}\,u^2 \,,
\ee
where the $K_m$ ($m=-1,0,+1$) are independent of $u$ and the $K_{\pm 1}$ are Hermitian. Decomposing $K_0$ into Hermitian and anti-Hermitian parts, $K_0 = L_{-1} -iL_{+1}$, we have
\be \label{general-D}
D =  K_{-1} \circ H_{-1} + L_{-1} \circ id + 2\,L_{+1}\circ H_0 + K_{+1}\circ H_{+1} \,,
\ee
where the symmetry generators
\be \label{sl(2)}
H_{-1} = i\,\partial_u \,,\quad H_0 = i\, u\,\partial_u - \frac{i}{2} \,,\quad H_{+1} = i\, u\,(u\,\partial_u - 1)
\ee
satisfy the $\mathfrak{sl}(2,\mathbb R)$ algebra (or $\mathfrak{gl}(2,\mathbb R)$ if one includes the identity), representing the conformal isometries of the real line.

The non-trivial higher symmetries of the action \eqref{eq:action} thus span a real Lie algebra isomorphic to
\be \label{higher-symm-algebra}
\cH(S^d) \otimes \mathfrak{gl}(2,\mathbb R) \,,
\ee
with $\cH(S^d)$ the Lie algebra of Hermitian differential operators on the celestial sphere $S^d$, while the set $\{id,H_{-1},H_0,H_{+1}\}$ forms a basis of $\mathfrak{gl}(2,\mathbb R)$. The commutator (times the imaginary unit~$i$) of differential operators defines a Lie bracket on this vector space.
As anticipated, this algebra is much bigger than $\mathfrak{ihs}_{d+2}$ and we now discuss some relevant subalgebras.

\subsection{Extended BMS symmetry and large \texorpdfstring{$\mathfrak{u}(1)$}{u(1)} transformations}\label{sec:extended-bms}

We begin by considering symmetries which are differential operators of order zero. They are real functions on the sphere $S^d$, $D = \alpha(x)$, corresponding to local phase transformations
\be
\delta_\alpha \varphi = i\,\alpha(x)\,\varphi \,.
\ee
In a putative holographic correspondence, this symmetry should signal the presence of large $\mathfrak{u}(1)$ transformations as in \cite{Strominger:2013lka} within the asymptotic symmetries of the bulk theory.

We now move to first-order symmetries and write $K_{-1} = -T(x)$ and $L_{-1} = -i\,Y^i(x)\,\partial_i - \frac{i}{2}\,\nabla_iY^i(x)$ with $T$ and $Y^i$ real, while choosing $L_{+1} = -\frac{1}{2d}\,\nabla_{\!i}Y^i(x)$. In this way, we obtain
\be \label{Dgbms}
i\,D_\mathfrak{ebms} = T(x)\,\partial_u + Y^i(x)\,\partial_i + \frac{1}{d}\,\nabla_{\!i} Y^i(x) \left(\Delta + u\,\partial_u\right) ,
\ee
which is the form of a differential operator of order one generating super-translations via $T(x)$ and super-rotations via $Y^i(x)$ when acting on a scalar density of scaling dimension $\Delta = \frac{d-1}{2}$. Note that we did not impose any constraint on the vectors $Y^i(x)$: therefore the super-rotations in \eqref{Dgbms} generate the whole $\mathfrak{diff}(S^d)$ algebra as in the extended BMS algebra of \cite{Campiglia:2014yka}.

\subsection{Enhanced BMS symmetry algebra \texorpdfstring{$\mathfrak{bms}^+_{d+2}$}{bms+}}\label{sec:mega-extension}

Still looking at first-order differential operators, there are two more available functions generating super-dilations and super-conformal boosts in the $u$ direction:
\be \label{general-1st-order}
D_\mathfrak{bms^+} = D_\mathfrak{ebms} + W(x)\,H_0 + Z(x)\,H_{+1} \,,
\ee
with real $W(x)$ and $Z(x)$.
From the bulk viewpoint, in which one seeks to interpret these transformations as being associated to asymptotic symmetries, the action of $W(x)\,H_0$ should correspond to ``BMS-Weyl'' transformations as those considered in \cite{Freidel:2021fxf}, while, to our knowledge, the action of $Z(x)\,H_{+1}$ has not been considered in the literature. We denote the full first-order subalgebra of \eqref{higher-symm-algebra} as $\mathfrak{bms}^+_{d+2}$ and we remark that it does not seem to be isomorphic to any of the proposed conformal extensions of the BMS algebra \cite{Haco:2017ekf, Fuentealba:2020zkf}.

\subsection{Algebra of higher-order operators}

The symmetrised product 
of differential operators satisfying \eqref{hermiticity} is a differential operator satisfying the same condition (similarly to the commutator times the imaginary unit, that defines the Lie bracket on \eqref{higher-symm-algebra}).
Higher-order symmetries can thus be realised as symmetrised products of first-order ones. This simple observation guarantees that $\mathfrak{ihs}_{d+2}$ is a Lie subalgebra of \eqref{higher-symm-algebra}. Indeed, as it is also manifest in \eqref{Dgbms}, the generators \eqref{realisation} belong to the symmetries of the simpleton and their symmetrised products acting on the simpleton give, by construction, the algebra $\mathfrak{ihs}_{d+2}$.

As discussed in section~\ref{sec:extended-bms}, the constraints \eqref{goodcut} and \eqref{conformalKilling} that select the algebra $\mathfrak{ihs}_{d+2}$ are not necessary to identify a symmetry of the simpleton. Therefore, one can also consider products of the operators in \eqref{Dgbms} with unconstrained $T(x)$ and $Y^i(x)$, 
which form a subalgebra. In this way one obtains an infinite-dimensional extension of the algebra $\mathfrak{ihs}_{d+2}$, that we dub $\mathfrak{hsbms}_{d+2}$. It corresponds to realising the $\mathfrak{ebms}_{d+2}$ UEA on the simpleton module, while the infinite-dimensional extension of \cite{Bekaert:2022ipg} realises it on the Sachs module. As a result, in that case as well as in \cite{Campoleoni:2017mbt, Campoleoni:2020ejn} polynomials of any order in $u$ appear in the differential operators, while here the $u$ dependence only comes from the operators \eqref{sl(2)}, compatibly with the observation that $\mathfrak{ihs}_{d+2}$ is not a subalgebra of the higher-spin algebras of \cite{Bekaert:2022ipg}.

Concretely, higher-order symmetries are obtained composing the operators \eqref{sl(2)} with higher-order differential operators on the sphere, see \eqref{higher-symm-algebra}. The latter can then be expanded, e.g., as
\be
L_{-1} = \sum_{s \geq 3} i^{s-1}\,Y^{j_1\cdots j_{s-1}} \,\nabla_{\!(j_1}\cdots\,\nabla_{\!j_{s-1})} + \text{lower derivative}
\ee
with real $Y^{j_1 \cdots j_{s-1}}$, and where lower-derivative terms are required to obtain Hermitian operators. The operators $K_{-1}$, $L_{+1}$ and $K_{+1}$ in \eqref{general-D} admit similar expansions. The tensors $Y^{j_1 \cdots j_{s-1}}$ are often called the symbol of their associated differential operator. We only need symmetric symbols because any antisymmetrisation can be reabsorbed into operators of lower order. 
A higher-spin generalisation of \mbox{(super-)}translations, i.e.\ an Abelian ideal, is given by $K_{-1}\circ H_{-1}$, while, in analogy with \eqref{Dgbms}, the remaining generators of $\mathfrak{hsbms}_{d+2}$ are given by linear combinations of $L_{-1} \circ id$ and $L_{+1}\circ H_0$,  
each one depending only on $Y^{i_1 \cdots i_{s-1}}$.
This symbol, in its turn, can be recovered as a symmetrised product of vectors $Y^i$. 
In spite of the differences discussed above, decomposing the traceful tensors $Y^{i_1 \cdots i_{s-1}}$ and the corresponding symbols $T^{i_1 \cdots i_{s-2}}$ of $K_{-1}$ into traceless parts, one recovers the same set of generators as in the asymptotic symmetries of Fronsdal fields for the weakest boundary conditions considered in \cite{Campoleoni:2020ejn}. 
We conclude by stressing that one can also consider products of the operators \eqref{general-1st-order}, so that the higher symmetries of the simpleton actually provide a higher-spin extension of the algebra $\mathfrak{bms}^+_{d+2}$ of section~\ref{sec:mega-extension}.

\section*{Acknowledgments}
We are grateful to B.~Oblak for discussions and collaboration at an early stage of this work. We thank I.~Basile, J.~Figueroa-O'Farrill, D.~Ponomarev and S.~Prohazka for useful discussions and X.B.\ thanks the University of Mons for hospitality. A.C.\ and S.P.\ are, respectively, a research associate and a FRIA grantee of the Fonds de la Recherche Scientifique -- FNRS. Their work was supported by FNRS under Grants No.\ FC.36447, F.4503.20 and T.0022.19. S.P.\ acknowledges the support of the SofinaBo\"el Fund for Education and Talent. This research was partially completed at the ``2${}^\text{nd}$ Carroll Workshop'', organised by the University of Mons, and at the workshop ``Higher Spin Gravity and its Applications'', supported by the Asia Pacific Center for Theoretical Physics.

%%%%%%%%%%%%%%%%%%%%%%%%%%%%%%%%%%%%%%%%%%%%%%%%%%%%%%%%%%%%%%%%%%%

%\bibliographystyle{my-elsarticle-num}
%\bibliographystyle{utphys}
%\bibliographystyle{JHEP}
%\bibliography{biblio.bib}

\begin{thebibliography}{10}
%\expandafter\ifx\csname url\endcsname\relax
%  \def\url#1{\texttt{#1}}\fi
%\expandafter\ifx\csname urlprefix\endcsname\relax\def\urlprefix{URL }\fi
%\expandafter\ifx\csname href\endcsname\relax
%  \def\href#1#2{#2} \def\path#1{#1}\fi

\bibitem{Pasterski:2021raf}
S.~Pasterski, M.~Pate and A.~M.~Raclariu,
\emph{Celestial Holography},
\href{https://arxiv.org/abs/2111.11392}{arXiv:2111.11392}.

\bibitem{Donnay:2022aba}
L.~Donnay, A.~Fiorucci, Y.~Herfray and R.~Ruzziconi,
\emph{Carrollian Perspective on Celestial Holography},
\href{https://doi.org/10.1103/PhysRevLett.129.071602}{Phys. Rev. Lett. \textbf{129} (2022) no.7, 071602}
%doi:10.1103/PhysRevLett.129.071602
[\href{https://arxiv.org/abs/2202.04702}{arXiv:2202.04702}].

\bibitem{Bagchi:2022emh}
A.~Bagchi, S.~Banerjee, R.~Basu and S.~Dutta,
\emph{Scattering Amplitudes: Celestial and Carrollian},
\href{https://doi.org/10.1103/PhysRevLett.128.241601}{Phys. Rev. Lett. \textbf{128} (2022) no.24, 241601}
%doi:10.1103/PhysRevLett.128.241601
[\href{https://arxiv.org/abs/2202.08438}{arXiv:2202.08438}].

\bibitem{Dappiaggi:2005ci}
C.~Dappiaggi, V.~Moretti and N.~Pinamonti,
\emph{Rigorous steps towards holography in asymptotically flat spacetimes},
\href{https://doi.org/10.1142/S0129055X0600270X}{Rev. Math. Phys. \textbf{18} (2006), 349-416}
%doi:10.1142/S0129055X0600270X
[\href{https://arxiv.org/abs/gr-qc/0506069}{arXiv:gr-qc/0506069}].

\bibitem{Kar:2022vqy}
A.~Kar, L.~Lamprou, C.~Marteau and F.~Rosso,
\emph{Celestial Matrix Model},
\href{https://doi.org/10.1103/PhysRevLett.129.201601}{Phys. Rev. Lett. \textbf{129} (2022) no.20, 201601}
%doi:10.1103/PhysRevLett.129.201601
[\href{https://arxiv.org/abs/2205.02240}{arXiv:2205.02240}].

\bibitem{Costello:2022jpg}
K.~Costello, N.~M.~Paquette and A.~Sharma,
\emph{Top-down holography in an asymptotically flat spacetime},
\href{https://arxiv.org/abs/2208.14233}{arXiv:2208.14233}.

\bibitem{Stieberger:2022zyk}
S.~Stieberger, T.~R.~Taylor and B.~Zhu,
\emph{Celestial Liouville Theory for Yang-Mills Amplitudes},
\href{https://doi.org/10.1016/j.physletb.2022.137588}{Phys. Lett. B \textbf{836} (2023), 137588}
%doi:10.1016/j.physletb.2022.137588
\href{https://arxiv.org/abs/2209.02724}{[arXiv:2209.02724]}.

\bibitem{Ponomarev:2022ryp}
D.~Ponomarev,
\emph{Towards higher-spin holography in flat space},
\href{https://arxiv.org/abs/2210.04035}{arXiv:2210.04035}.

\bibitem{Bekaert:2012ux}
X.~Bekaert, E.~Joung and J.~Mourad,
\emph{Comments on higher-spin holography},
\href{https://doi.org/10.1002/prop.201200014}{Fortsch. Phys. \textbf{60} (2012), 882-888}
%doi:10.1002/prop.201200014
[\href{https://arxiv.org/abs/1202.0543}{arXiv:1202.0543}].

\bibitem{Giombi:2016ejx}
S.~Giombi,
\emph{Higher Spin \textemdash{} CFT Duality},
%doi:10.1142/9789813149441\_0003
\href{https://arxiv.org/abs/1607.02967}{arXiv:1607.02967}.

\bibitem{Dirac:1963ta}
P.~A.~M.~Dirac,
\emph{A Remarkable representation of the 3 + 2 de Sitter group},
\href{https://doi.org/10.1063/1.1704016}{J. Math. Phys. \textbf{4} (1963), 901-909}.
%doi:10.1063/1.1704016

\bibitem{Flato:1978qz}
M.~Flato and C.~Fronsdal,
\emph{One Massless Particle Equals Two Dirac Singletons: Elementary Particles in a Curved Space. 6.},
\href{https://doi.org/10.1007/BF00400170}{Lett. Math. Phys. \textbf{2} (1978), 421-426}.
%doi:10.1007/BF00400170

\bibitem{Vasiliev:2004cm}
M.~A.~Vasiliev,
\emph{Higher spin superalgebras in any dimension and their representations},
\href{https://doi.org/10.1088/1126-6708/2004/12/046}{JHEP \textbf{12} (2004), 046}
%doi:10.1088/1126-6708/2004/12/046
[\href{https://arxiv.org/abs/hep-th/0404124}{arXiv:hep-th/0404124}].

\bibitem{Vasiliev:2003ev}
M.~A.~Vasiliev,
\emph{Nonlinear equations for symmetric massless higher spin fields in (A)dS(d)},
\href{https://doi.org/10.1016/S0370-2693(03)00872-4}{Phys. Lett. B \textbf{567} (2003), 139-151}
%doi:10.1016/S0370-2693(03)00872-4
[\href{https://arxiv.org/abs/hep-th/0304049}{arXiv:hep-th/0304049}].

\bibitem{Vasiliev:1999ba}
M.~A.~Vasiliev,
\emph{Higher spin gauge theories: Star product and AdS space},
%doi:10.1142/9789812793850\_0030
\href{https://arxiv.org/abs/hep-th/9910096}{arXiv:hep-th/9910096}.

\bibitem{Eastwood:2002su}
M.~G.~Eastwood,
\emph{Higher symmetries of the Laplacian},
\href{https://doi.org/10.4007/annals.2005.161.1645}{Annals Math. \textbf{161} (2005), 1645-1665}
%doi:10.4007/annals.2005.161.1645
[\href{https://arxiv.org/abs/hep-th/0206233}{arXiv:hep-th/0206233}].

\bibitem{Iazeolla:2008ix}
C.~Iazeolla and P.~Sundell,
\emph{A Fiber Approach to Harmonic Analysis of Unfolded Higher-Spin Field Equations},
\href{https://doi.org/10.1088/1126-6708/2008/10/022}{JHEP \textbf{10} (2008), 022}
%doi:10.1088/1126-6708/2008/10/022
[\href{https://arxiv.org/abs/0806.1942}{arXiv:0806.1942}].

\bibitem{Campoleoni:2021blr}
A.~Campoleoni and S.~Pekar,
\emph{Carrollian and Galilean conformal higher-spin algebras in any dimensions},
\href{https://doi.org/10.1007/JHEP02(2022)150}{JHEP \textbf{02} (2022), 150}
%doi:10.1007/JHEP02(2022)150
[\href{https://arxiv.org/abs/2110.07794}{arXiv:2110.07794}].

\bibitem{Bagchi:2019xfx}
A.~Bagchi, A.~Mehra and P.~Nandi,
\emph{Field Theories with Conformal Carrollian Symmetry},
\href{https://doi.org/10.1007/JHEP05(2019)108}{JHEP \textbf{05} (2019), 108}
%doi:10.1007/JHEP05(2019)108
[\href{https://arxiv.org/abs/1901.10147}{arXiv:1901.10147}].

\bibitem{Henneaux:2021yzg}
M.~Henneaux and P.~Salgado-Rebolledo,
\emph{Carroll contractions of Lorentz-invariant theories},
\href{https://doi.org/10.1007/JHEP11(2021)180}{JHEP \textbf{11} (2021), 180}
%doi:10.1007/JHEP11(2021)180
[\href{https://arxiv.org/abs/2109.06708}{arXiv:2109.06708}].

\bibitem{deBoer:2021jej}
J.~de Boer, J.~Hartong, N.~A.~Obers, W.~Sybesma and S.~Vandoren,
\emph{Carroll Symmetry, Dark Energy and Inflation},
\href{https://doi.org/10.3389/fphy.2022.810405}{Front. in Phys. \textbf{10} (2022), 810405}
%doi:10.3389/fphy.2022.810405
[\href{https://arxiv.org/abs/2110.02319}{arXiv:2110.02319}].

\bibitem{Gupta:2020dtl}
N.~Gupta and N.~V.~Suryanarayana,
\emph{Constructing Carrollian CFTs},
\href{https://doi.org/10.1007/JHEP03(2021)194}{JHEP \textbf{03} (2021), 194}
%doi:10.1007/JHEP03(2021)194
[\href{https://arxiv.org/abs/2001.03056}{arXiv:2001.03056}].

\bibitem{Rivera-Betancour:2022lkc}
D.~Rivera-Betancour and M.~Vilatte,
\emph{Revisiting the Carrollian scalar field},
\href{https://doi.org/10.1103/PhysRevD.106.085004}{Phys. Rev. D \textbf{106} (2022) no.8, 085004}
%doi:10.1103/PhysRevD.106.085004
[\href{https://arxiv.org/abs/2207.01647}{arXiv:2207.01647}].

\bibitem{Baiguera:2022lsw}
S.~Baiguera, G.~Oling, W.~Sybesma and B.~T.~S\o{}gaard,
\emph{Conformal Carroll Scalars with Boosts},
\href{https://arxiv.org/abs/2207.03468}{arXiv:2207.03468}.

\bibitem{Duval:2014uva}
C.~Duval, G.~W.~Gibbons and P.~A.~Horvathy,
\emph{Conformal Carroll groups and BMS symmetry},
\href{https://doi.org/10.1088/0264-9381/31/9/092001}{Class. Quant. Grav. \textbf{31} (2014), 092001}
%doi:10.1088/0264-9381/31/9/092001
[\href{https://arxiv.org/abs/1402.5894}{arXiv:1402.5894}].

\bibitem{Nikitin1991-gp}
A.G.~Nikitin,
\emph{Generalized killing tensors of arbitrary rank and order},
%doi:10.1007/BF01058941
\href{https://doi.org/10.1007/BF01058941}{Ukr. Math. J. \textbf{43} (1991), 734–743}.

\bibitem{Arcioni:2003td}
G.~Arcioni and C.~Dappiaggi,
\emph{Holography in asymptotically flat space-times and the BMS group},
\href{https://doi.org/10.1088/0264-9381/21/23/022}{Class. Quant. Grav. \textbf{21} (2004), 5655}
%doi:10.1088/0264-9381/21/23/022
[\href{https://arxiv.org/abs/hep-th/0312186}{arXiv:hep-th/0312186}].

\bibitem{Bagchi:2016bcd}
A.~Bagchi, R.~Basu, A.~Kakkar and A.~Mehra,
\emph{Flat Holography: Aspects of the dual field theory},
\href{https://doi.org/10.1007/JHEP12(2016)147}{JHEP \textbf{12} (2016), 147}
%doi:10.1007/JHEP12(2016)147
[\href{https://arxiv.org/abs/1609.06203}{arXiv:1609.06203}].

\bibitem{Ciambelli:2018wre}
L.~Ciambelli, C.~Marteau, A.~C.~Petkou, P.~M.~Petropoulos and K.~Siampos,
\emph{Flat holography and Carrollian fluids},
\href{https://doi.org/10.1007/JHEP07(2018)165}{JHEP \textbf{07} (2018), 165}
%doi:10.1007/JHEP07(2018)165
[\href{https://arxiv.org/abs/1802.06809}{arXiv:1802.06809}].

\bibitem{Ponomarev:2022qkx}
D.~Ponomarev,
\emph{Chiral higher-spin holography in flat space: the Flato-Fronsdal theorem and lower-point functions},
\href{https://arxiv.org/abs/2210.04036}{arXiv:2210.04036}.

\bibitem{Ponomarev:2016lrm}
D.~Ponomarev and E.~D.~Skvortsov,
\emph{Light-Front Higher-Spin Theories in Flat Space},
\href{https://doi.org/10.1088/1751-8121/aa56e7}{J. Phys. A \textbf{50} (2017) no.9, 095401}
%doi:10.1088/1751-8121/aa56e7
[\href{https://arxiv.org/abs/1609.04655}{arXiv:1609.04655}].

\bibitem{Bekaert:2008sa}
X.~Bekaert,
\emph{Comments on higher-spin symmetries},
\href{https://doi.org/10.1142/S0219887809003527}{Int. J. Geom. Meth. Mod. Phys. \textbf{6} (2009), 285-342}
%doi:10.1142/S0219887809003527
[\href{https://arxiv.org/abs/0807.4223}{arXiv:0807.4223}].

\bibitem{Bekaert:2022ipg}
X.~Bekaert and B.~Oblak,
\emph{Massless Scalars and Higher-Spin BMS in Any Dimension},
\href{https://doi.org/10.1007/JHEP11(2022)022}{JHEP \textbf{11} (2022), 022}
%doi:10.1007/JHEP11(2022)022
[\href{https://arxiv.org/abs/2209.02253}{arXiv:2209.02253}].

\bibitem{Campoleoni:2017mbt}
A.~Campoleoni, D.~Francia and C.~Heissenberg,
\emph{On higher-spin supertranslations and superrotations},
\href{https://doi.org/10.1007/JHEP05(2017)120}{JHEP \textbf{05} (2017), 120}
%doi:10.1007/JHEP05(2017)120
[\href{https://arxiv.org/abs/1703.01351}{arXiv:1703.01351}].

\bibitem{Campoleoni:2020ejn}
A.~Campoleoni, D.~Francia and C.~Heissenberg,
\emph{On asymptotic symmetries in higher dimensions for any spin},
\href{https://doi.org/10.1007/JHEP12(2020)129}{JHEP \textbf{12} (2020), 129}
%doi:10.1007/JHEP12(2020)129
[\href{https://arxiv.org/abs/2011.04420}{arXiv:2011.04420}].

\bibitem{Ammon:2020fxs}
M.~Ammon, M.~Pannier and M.~Riegler,
\emph{Scalar Fields in 3D Asymptotically Flat Higher-Spin Gravity},
\href{https://doi.org/10.1088/1751-8121/abdbc6}{J. Phys. A \textbf{54} (2021) no.10, 105401}
%doi:10.1088/1751-8121/abdbc6
[\href{https://arxiv.org/abs/2009.14210}{arXiv:2009.14210}].

\bibitem{Joung:2015jza}
E.~Joung and K.~Mkrtchyan,
\emph{Partially-massless higher-spin algebras and their finite-dimensional truncations},
\href{https://doi.org/10.1007/JHEP01(2016)003}{JHEP \textbf{01} (2016), 003}
%doi:10.1007/JHEP01(2016)003
[\href{https/arxiv.org/abs/1508.07332}{arXiv:1508.07332}].

\bibitem{Ponomarev:2021xdq}
D.~Ponomarev,
\emph{3d conformal fields with manifest sl(2, \ensuremath{\mathbb{C}})},
\href{https://doi.org/10.1007/JHEP06(2021)055}{JHEP \textbf{06} (2021), 055}
%doi:10.1007/JHEP06(2021)055
[\href{https://arxiv.org/abs/2104.02770}{arXiv:2104.02770}].

\bibitem{Herfray:2021qmp}
Y.~Herfray,
\emph{Carrollian manifolds and null infinity: a view from Cartan geometry},
\href{https://doi.org/10.1088/1361-6382/ac635f}{Class. Quant. Grav. \textbf{39} (2022) no.21, 215005}
%doi:10.1088/1361-6382/ac635f
[\href{https://arxiv.org/abs/2112.09048}{arXiv:2112.09048}].

\bibitem{Campiglia:2014yka}
M.~Campiglia and A.~Laddha,
\emph{Asymptotic symmetries and subleading soft graviton theorem},
\href{https://doi.org/10.1103/PhysRevD.90.124028}{Phys. Rev. D \textbf{90} (2014) no.12, 124028}
%doi:10.1103/PhysRevD.90.124028
[\href{https://arxiv.org/abs/1408.2228}{arXiv:1408.2228}].

\bibitem{Bekaert:2012vt}
X.~Bekaert and M.~Grigoriev,
\emph{Notes on the ambient approach to boundary values of AdS gauge fields},
\href{https://doi.org/10.1088/1751-8113/46/21/214008}{J. Phys. A \textbf{46} (2013), 214008}
%doi:10.1088/1751-8113/46/21/214008
[\href{https://arxiv.org/abs/1207.3439}{arXiv:1207.3439}].

\bibitem{Herfray:2020rvq}
Y.~Herfray,
\emph{Asymptotic shear and the intrinsic conformal geometry of null-infinity},
\href{https://doi.org/10.1063/5.0003616}{J. Math. Phys. \textbf{61} (2020) no.7, 072502}
%doi:10.1063/5.0003616
[\href{https://arxiv.org/abs/2001.01281}{arXiv:2001.01281}].

\bibitem{Figueroa-OFarrill:2018ilb}
J.~Figueroa-O'Farrill and S.~Prohazka,
\emph{Spatially isotropic homogeneous spacetimes},
\href{https://doi.org/10.1007/JHEP01(2019)229}{JHEP \textbf{01} (2019), 229}
%doi:10.1007/JHEP01(2019)229
[\href{https://arxiv.org/abs/1809.01224}{arXiv:1809.01224}].

\bibitem{Bekaert:2013zya}
X.~Bekaert and M.~Grigoriev,
\emph{Higher order singletons, partially massless fields and their boundary values in the ambient approach},
\href{https://doi.org/10.1016/j.nuclphysb.2013.08.015}{Nucl. Phys. B \textbf{876} (2013), 667-714}
%doi:10.1016/j.nuclphysb.2013.08.015
[\href{https://arxiv.org/abs/1305.0162}{arXiv:1305.0162}].

\bibitem{Shapovalov1992-tf}
A.V.~Shapovalov and I.V.~Shirokov,
\emph{Symmetry algebras of linear differential equations},
%doi:10.1007/BF01018697
\href{https://doi.org/10.1007/BF01018697}{Theor. Math. Phys. 92 (1992), 697–703.}

\bibitem{Bidussi:2021nmp}
L.~Bidussi, J.~Hartong, E.~Have, J.~Musaeus and S.~Prohazka,
\emph{Fractons, dipole symmetries and curved spacetime},
\href{https://doi.org/10.21468/SciPostPhys.12.6.205}{SciPost Phys. \textbf{12} (2022) no.6, 205}
%doi:10.21468/SciPostPhys.12.6.205
[\href{https://arxiv.org/abs/2111.03668}{arXiv:2111.03668}].

\bibitem{Hao:2021urq}
P.~x.~Hao, W.~Song, X.~Xie and Y.~Zhong,
\emph{BMS-invariant free scalar model},
\href{https://doi.org/10.1103/PhysRevD.105.125005}{Phys. Rev. D \textbf{105} (2022) no.12, 125005}
%doi:10.1103/PhysRevD.105.125005
[\href{https://arxiv.org/abs/1308.0589}{arXiv:2111.04701}].

\bibitem{Bagchi:2022eav}
A.~Bagchi, A.~Banerjee, S.~Dutta, K.~S.~Kolekar and P.~Sharma,
\emph{Carroll covariant scalar fields in two dimensions},
\href{https://arxiv.org/abs/2203.13197}{arXiv:2203.13197}.

\bibitem{Strominger:2013lka}
A.~Strominger,
\emph{Asymptotic Symmetries of Yang-Mills Theory},
\href{https://doi.org/10.1007/JHEP07(2014)151}{JHEP \textbf{07} (2014), 151}
%doi:10.1007/JHEP07(2014)151
[\href{https://arxiv.org/abs/1308.0589}{arXiv:1308.0589}].

\bibitem{Freidel:2021fxf}
L.~Freidel, R.~Oliveri, D.~Pranzetti and S.~Speziale,
\emph{The Weyl BMS group and Einstein\textquoteright{}s equations},
\href{https://doi.org/10.1007/JHEP07(2021)170}{JHEP \textbf{07} (2021), 170}
%doi:10.1007/JHEP07(2021)170
[\href{https://arxiv.org/abs/2104.05793}{arXiv:2104.05793}].

\bibitem{Haco:2017ekf}
S.~J.~Haco, S.~W.~Hawking, M.~J.~Perry and J.~L.~Bourjaily,
\emph{The Conformal BMS Group},
\href{https://doi.org/10.1007/JHEP11(2017)012}{JHEP \textbf{11} (2017), 012}
%doi:10.1007/JHEP11(2017)012
[\href{https://arxiv.org/abs/1701.08110}{arXiv:1701.08110}].

\bibitem{Fuentealba:2020zkf}
O.~Fuentealba, H.~A.~Gonz\'alez, A.~P\'erez, D.~Tempo and R.~Troncoso,
\emph{Superconformal Bondi-Metzner-Sachs Algebra in Three Dimensions},
\href{https://doi.org/10.1103/PhysRevLett.126.091602}{Phys. Rev. Lett. \textbf{126} (2021) no.9, 091602}
%doi:10.1103/PhysRevLett.126.091602
[\href{https://arxiv.org/abs/2011.08197}{arXiv:2011.08197}].

\end{thebibliography}


%\begin{thebibliography}{00}


%\bibitem{}

%\end{thebibliography}
\end{document}